# A NEW SUBBAND NON LINEAR PREDICTION CODING ALGORITHM FOR NARROWBAND SPEECH SIGNAL: THE *n*ADPCMB⊥MLT CODING SCHEME

*Guido D'Alessandro[*], Marcos Faundez Zanuy[**], Francesco Piazza[*]*

[*]Dep. of Electronics and Automatics, University of Ancona, Ancona, Italy
e-mail: gzdedo@wolfpc.ea.unian.it, upf@ieee.org; http://eleweb.ea.unian.it
[**]Dep. of Telecommunications, Escola Universitària Politècnica de Mataró, Barcelona, Spain
e-mail: faundez@eupmt.es; http://www.eupmt.es/veu

**ABSTRACT**

This paper focuses on a newly developed transparent *n*ADPCMB⊥MLT speech coding algorithm. Our coder first decomposes the narrowband speech signal in subbands, a non linear ADPCM scheme is then performed in each subband. The signal subband decomposition is piloted by the equivalent Modulated Lapped Transform (MLT) filter bank. The novelty of this algorithm is the non linear approach, based on neural networks, to subband prediction coding. We have evaluated the performance of the *n*ADPCMB⊥MLT's coding algorithm with a session of formal listening based on the five grade impairment scale standardized within ITU – T Recommendation P.800.

## 1. INTRODUCTION

The frequency domain approach to coding represents the actual state of the art in speech and multi - purpose audio coding. In the non linear context the proposed algorithm will therefore use this approach.

After some experiments with Discrete Cosine Transform (DCT) and Modulated Lapped Transform (MLT) [1] we decided to investigate only non linear prediction in the subbands built with Modulated Lapped Transform, since with DCT the reconstructed signal may have discontinuities at block edge. This drawback is recognizable like "clicks" or "burbles" at the block frequency $F_s/M$ [3] and their unwished effects can stress the subband quantizers and the subband predictors. We have considered a number of uniform subbands, M, equal to 16. This number assures to treat a narrowband speech signal with a frequency resolution of 250 Hz. Each of the sixteen subbands covers, more or less, two critical bands, in the Bark scale, under 2 kHz, namely where the speech frequency formants are located. Hence this frequency resolution should be a good choice to code a narrowband speech signal, that frequently is a poor signal. Besides we found out that with this number of frequency components, or bands, we reached a fair compromise between frequency and temporal resolution in this application.

According to the definition of MLT transform [1] we used the half sine window as a low pass filter modulated by the MLT. That choice assures, in absence of subband coding, a perfect reconstruction of the signal after having applied the Inverse Modulated Lapped Transform (IMLT), or backward MLT [1], for signal conversion from subband representation to time representation.

By exploiting time correlation among samples of the same frequency component in each subband, we used a non linear prediction to predict the next value of frequency *bins*. In this sense we put in each subband a non linear ADPCM backward scheme ( *n*ADPCMB, see Fig.1 in the next page ), based on neural nets, similar to the type proposed in [5].

We have evaluated our novel coding algorithm with a formal listening test, observing the MOS ( *Mean Opinion Score* ) guideline, for the following bit rates: 32 kbps and 24 kbps. These bit rates have only been used to evaluate the feasibility of the idea. Other more efficient bit allocation algorithms and analysis filter banks could be used in future works to achieve more interesting compression rates within the non linear subband prediction algorithm. Considering, as we do in the second section, that the used quantizer isn't optimized to work with non linear predictor, neither with the statistics of subband signals, the obtained results are hopeful. In the fourth section, among the open problem, we will propose solutions to the subband quantizer optimization.

## 2. THE *n*ADPCMB⊥MLT CODING ALGORITHM

Our idea consists of a mix of two different types of coding: 1) subband coding and 2) non linear prediction coding. In order to *realize* this idea we implemented the *n*ADPCMB⊥MLT coding algorithm; his block scheme is proposed in figure 2. The acronym *n*ADPCMB⊥MLT suggests very well the basic components of the novel idea and the symbol of orthogonality means that non linear ADPCM works orthogonally to the filter bank.

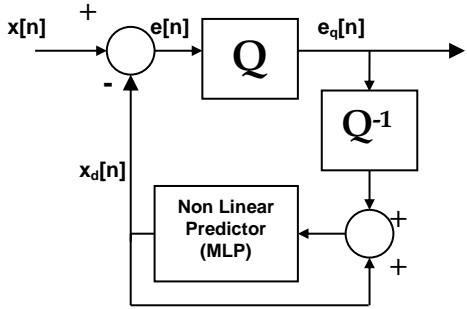

Fig. 1 ADPCM coder structure, backward configuration.

By Fig.1 it can be deduced that, since we have implemented the non linear ADPCM scheme in the backward configuration [3], it is possible to recover the decoded samples of each subband signal also at the transmitter side ( by the signal $x_d$ ).

With the proposed *n*ADPCMB⊥MLT scheme ( see Fig.2 ) our intention is to discover the possibility of using the non linear approach, proposed in [5], also in subband coding and, more in general, in hybrid frequency coding. Clearly there is one basic problem to investigate: the knowledge of time correlation between temporally successive *frequency bins*. In order to examine in depth this crucial problem we have tested the *n*ADPCMB⊥MLT algorithm using different lengths of the *bin* frames during the non linear prediction training. Approaching the problem in this way is possible understanding how to change the optimization of the network structure parameters reached in [5].

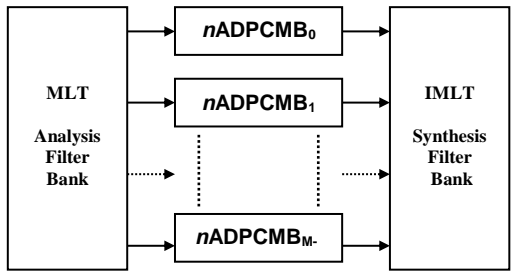

Fig. 2 *n*ADPCMB⊥MLT coder structure.

## 2.1 Algorithm overview
*Subbands predictors coefficients updating*:

- The coefficients are updated once every *bin* frame.

- An ADPCM backward configuration [3] is adopted to avoid the transmission of predictor coefficients, that are weights and biases of the network. Predictor coefficients are computed over the previous decoded *bin* frame, since it is already available at the receiver, that hence can compute the same values of coefficients without any additional information.

- The non linear analysis structure in each subband consists in a Multi - Layer Perceptron [4], that essentially derives from [5], with 10 input neurons, 2 hidden neurons with a sigmoid activation function and 1 output neuron with a linear activation function. The selected network architecture is trained with the Levenberg – Marquardt algorithm [6], that is one of the best optimizations of the Back Propagation algorithm; it's characterized by a fast convergence and a great reduction of necessary number of epochs.

*Residual subband prediction error quantization:*

- The residual prediction error in each subband has been quantized with the fixed bit allocation scheme shown in Table 1 for 32 kbps bit rate and in Table 2 for 24 kbps bit rate.

- The quantization step used in each subband is adapted by multiplier factors, obtained from [3]; $\Delta_{max}$ and $\Delta_{min}$ were set empirically.

| band | 1 | 2 | 3 | 4 | 5 | 6 | 7 | 8 | 9 | 10 | 11 | 12 | 13 | 14 | 15 | 16 |
|---|---|---|---|---|---|---|---|---|---|---|---|---|---|---|---|---|
| bits | 5 | 5 | 5 | 5 | 5 | 5 | 5 | 5 | 4 | 4 | 4 | 4 | 2 | 2 | 2 | 2 |

Table 1. Bit Allocation scheme of the proposed 24 kbps *n*ADPCMB⊥MLT algorithm.

| band | 1 | 2 | 3 | 4 | 5 | 6 | 7 | 8 | 9 | 10 | 11 | 12 | 13 | 14 | 15 | 16 |
|---|---|---|---|---|---|---|---|---|---|---|---|---|---|---|---|---|
| bits | 5 | 5 | 4 | 4 | 3 | 3 | 3 | 3 | 3 | 3 | 3 | 3 | 2 | 2 | 2 | 2 |

Table 2. Bit Allocation scheme of the proposed 24 kbps *n*ADPCMB⊥MLT algorithm.

## 2.2 Training parameter selection

Working with non linear prediction schemes based on neural nets, we have a number of parameters that must be optimized. We have found out that the performance of the non linear predictor into the closed loop ADPCM scheme, put in each subband (*n*ADPCMB of Fig. 1), are very sensible to the setting of:

- *Number of trained epochs*. In order to encode a frame, given by temporally subsequent *bins*, the neural network is trained with the previous *bin* frame, since we are in presence of a backward configuration. In order to obtain a good generalization capability treating inputs never *seen* during the training phase, it is important to try avoiding the over training problem. This is a crucial aspect for our algorithm, being based on a closed loop structure. In fact the signal used to train the network is corrupted by the quantization noise. The fewest quantization bits are, the most important the problem is. Considering this aspect, we have designed a fixed bit allocation scheme so as to assign few bits ( in number of two ) only to the high frequency bands, namely in our application at 32 kbps rate , over 3 kHz. ( It is known that often these frequency band are not very important to recover a good perceived quality. ) In this way we have preserved the frequency formants at 32 kbps rate, achieving a good and, in some cases, transparent quality with a bit rate of 32 kbps against the 128 kbps of the original signals. At 24 kbps rate the fixed bit allocation is more roughly and the MOS results of Tables 5 and 6 confirm it. In order to make the neural nets and thus our *n*ADPCMB⊥MLT algorithm as robust as possible to variations caused by the quantization noise, optimization of training conditions, such as number of epochs, used for the training, and number of random initialization of weights and biases has been performed.

- *Efficient initialization*. We used a multi start algorithm to achieve a good initialization, consisting in computing several Nguyen – Widrow initialization (experimentally fixed to 4) and choosing the one accomplishing the best validation. Considering that sometimes subsequent *bin* frames could be very similar, and so will do their associated weights, we substituted one of the random initialization with a deterministic initialization, using the weights of the previous frame. We have observed that this deterministic initialization can be trained less than the random one, since it is more close to the optimal solution. The Nguyen –Widrow initialization chooses initial values such that the active regions of the layer's units are roughly evenly distributed over the range of the input vector space. The benefit of using this method is that fewer units are wasted and that the network converges faster compared to purely random initial values.

A cross - validation procedure was used to find the right number of epochs and to control the multi start algorithm during the training process. The use of this procedure is necessary since it is not possible to verify, in a subband coding scheme, the level of training across one *bin* frame by the back control of the segSNR in the same *bin* frame similarly to the procedure that was proposed in [5]. Thus, the best epoch number to use during the training and the best initialization are both controlled and selected by the cross – validation procedure. In this sense our scheme implements a total backward configuration. Moreover the validation procedure assures to obtain an opportune level of training that allows a good generalization during prediction. An example of the results obtained by the cross –validation frame across one bins frame are reported in the Fig. 3. For the cross - validation procedure we have considered half of the available data for the training, one fourth for the validation and one fourth for the test. The maximum number of epochs was experimentally fixed to 50.

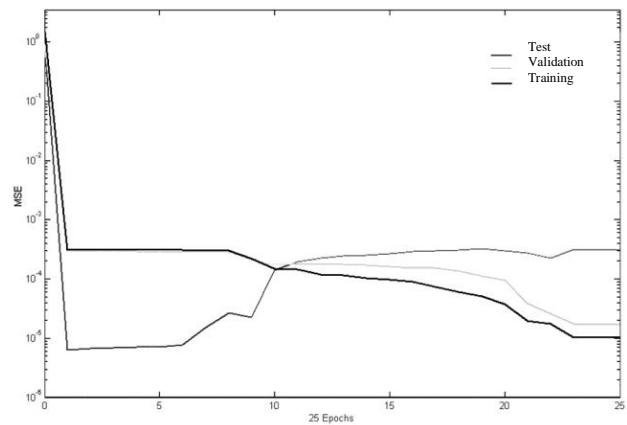

Fig.3 Results of cross - validation procedure across one *bins* frame.

### 3. EVALUATION AND EXPERIMENTAL RESULTS

The quality of the reconstructed speech samples was evaluated by a number of 20 subjects that carried out a formal listening test. The type of test is reported in the ITU Recommendation P.800 [2].

The listeners were presented the coded speech samples in a random order using a professional headphone, so that the perceived quality wasn't affected by the environmental and simulation system (PC) noises. The tables in the following page ( see tables 1, 2, 3, 4 ) show the MOS obtained by *n*ADPCMB⊥MLT at the bit rate of 24 and 32 kbps. At 24 kbps rate the quantization of some bands with 2 bits produces a strong reduction of the reconstructed signal quality. With 2 bits of quantization the decoded *bin* samples used by the network to train itself and to predict the next subband signals are very damaged by the quantization noise. In this sense 32 kbps bit allocation scheme of Table 1 and hence the training frames are more protected from the quantization noise.

### 3.1. Characteristics of the used speech signal database

Sixteen short narrowband speech signals (meanly long 2 or 4 seconds) were coded using the new *n*ADPCMB⊥MLT coder at the rate of 32 - kbit/s and 24 kbit/s. Each sentence was originally recorded using 16 quantization bits for sample at a sampling frequency of 8 kHz ( 16 kHz ), in other words a rate of 128 kbit/s. We extract these sixteen recordings from the DARPA TI MIT database having attention to choose 4 female speakers and 4 male ones, each of them uttering the same two sentences. The choice of use this high quality database in this work was done to assure compatibility between our tests and other works in the field of narrowband and wideband speech coding research.

| MOS | MCPM0SA1 | MARC0SA1 | MADC0SA1 | MARW0SA1 |
|---|---|---|---|---|
| 32kbps | 4.8 | 4.7 | 4.3 | 4.7 |
| MOS | MCPM0SA2 | MARC0SA2 | MADC0SA2 | MARW0SA2 |
| 32kbps | 4.6 | 4.7 | 4.1 | 4.7 |

Table 3. MOS results obtained for male 32 kbps *n*ADPCMB⊥MLT coded signals.

| MOS | FAKS0SA1 | FAJW0SA1 | FALK0SA1 | FDG0SA1 |
|---|---|---|---|---|
| 32kbps | 4.7 | 4.9 | 4.2 | 4.8 |
| MOS | FAKS0SA2 | FAJW0SA2 | FALK0SA2 | FDG0SA2 |
| 32kbps | 4.3 | 4.8 | 4.1 | 4.7 |

Table 4. MOS results obtained for female 32 kbps *n*ADPCMB⊥MLT coded signals.

| MOS | MCPM0SA1 | MARC0SA1 | MADC0SA1 | MARW0SA1 |
|---|---|---|---|---|
| 24kbps | 4.3 | 3.9 | 3.3 | 3.7 |
| MOS | MCPM0SA2 | MARC0SA2 | MADC0SA2 | MARW0SA2 |
| 24kbps | 4.1 | 3.6 | 3.1 | 3.2 |

Table 5. MOS results obtained for male 24 kbps *n*ADPCMB⊥MLT coded signals.

| MOS | FAKS0SA1 | FAJW0SA1 | FALK0SA1 | FDG0SA1 |
|---|---|---|---|---|
| 24kbps | 3.2 | 3.5 | 3.2 | 3.8 |
| MOS | FAKS0SA2 | FAJW0SA2 | FALK0SA2 | FDG0SA2 |
| 24kbps | 3.1 | 3.2 | 2.9 | 3.9 |

Table 6. MOS results obtained for female 24 kbps *n*ADPCMB⊥MLT coded signals.

## 4. CONCLUSIONS AND OPEN PROBLEMS

This work started with the observation that there are no works that deals in general terms with non linear prediction of speech signal in the frequency domain, for this reason we hope that our work could address and stimulate the research in this direction.

We have investigated the performance of *n*ADPCMB⊥MLT algorithm. The use of non linear subbands analysis leads to a good subbands prediction coding like it is confirmed by the above reported MOS tables. The problems of this algorithm reside in the difficult of a proper setting of all the parameters that condition the correct working of the algorithm. Above all we have observed its dependence on the number of initialization, in fact increasing their number the listening quality of the reconstructed signal improves, but at the cost of a major computational burden.

The *n*ADPCMB⊥MLT novel non linear scheme has been always stable in our experiments. The results with the non linear subbands coding of narrowband speech signals are hopeful.

The open problems concern essentially the subband quantizers and have, from our point of view, three possible solutions: 1) the optimization of the subband quantizers in order to work well with non linear subband predictors; 2) the optimization of quantizers to adapt its characteristic to the dynamic of subband signals; 3) changing quantizers typology is possible to develop a dynamic bits allocation algorithm, like the one used in the state of the arts frequency domain coders that could help to reduce the bit rate.

## 5. ACKNOWLEDGMENTS

We would like to express our gratitude to Prof. Bastiaan Kleijn, IEEE fellow member, for his useful comments about the right procedure to evaluate the quality of our new coder and to Ing. Rico Malvar for his helpfulness during the first steps of the work.

This work has been partially supported by EEUU SOCRATES ERASMUS program and by the CICYT TIC2000-1669-C04-02.